\documentclass[conference]{IEEEtran}
\usepackage{graphicx,
	psfrag,
	epsfig,
	epstopdf,
	amsthm,
    amsmath,
	amssymb,
	url,
	subfigure,
	algorithm,
	algorithmic,
	balance,
	enumerate,
	color,
    url,
	setspace,
    placeins,
}
\usepackage{blindtext}
\usepackage{cite}
\usepackage{microtype}

\newcommand{\cref}[1]{Constraint~\ref{#1}}

\newcommand{\ignore}[1]{}

\begin{document}
\title{Cost-optimal V2X Service Placement in Distributed Cloud/Edge Environment}	
%
\author{
\IEEEauthorblockN{Abdallah Moubayed\IEEEauthorrefmark{1}, Abdallah Shami\IEEEauthorrefmark{1}, Parisa Heidari\IEEEauthorrefmark{2}, Adel Larabi\IEEEauthorrefmark{3}, and Richard Brunner\IEEEauthorrefmark{2}} 
\IEEEauthorblockA{\IEEEauthorrefmark{1} Western University, London, Ontario, Canada; e-mails: \{amoubaye, abdallah.shami\}@uwo.ca\\
}
\IEEEauthorblockA{\IEEEauthorrefmark{2}  Ericsson, Montreal, Quebec, Canada; e-mails: \{parisa.heidari, richard.brunner\}@ericsson.com}
\IEEEauthorblockA{\IEEEauthorrefmark{3}  Edge Gravity by Ericsson, Montreal, Quebec, Canada; e-mail: ALarabi@edgegravity.ericsson.com}	
}
\maketitle
\begin{abstract}
	Deploying V2X services has become a challenging task. This is mainly due to the fact that such services have strict latency requirements. To meet these requirements, one potential solution is adopting mobile edge computing (MEC). However, this presents new challenges including how to find a cost efficient placement that meets other requirements such as latency. In this work, the problem of cost-optimal V2X service placement (CO-VSP) in a distributed cloud/edge environment is formulated. Additionally, a cost-focused delay-aware V2X service placement (DA-VSP) heuristic algorithm is proposed. Simulation results show that both CO-VSP model and DA-VSP algorithm guarantee the QoS requirements of all such services and illustrates the trade-off between latency and deployment cost.
\end{abstract}
\section{Introduction}\label{intro}
\indent Vehicle-to-everything (V2X) communication, a cornerstone of intelligent transportation systems (ITSs), has garnered more attention from different entities including governmental agencies and automotive manufacturers due to the various projected merits of such systems. For example, such systems help reduce road accidents, introduce new business opportunities, and reduce the cost of managing vehicular fleets  \cite{v2x_access_survey}. Additionally, V2X communication helps support new set of services such as traffic optimization and in-car entertainment \cite{v2x_communication_modes}. However, offering such services is coupled with maintaining stringent performance requirements. More specifically, end-to-end (E2E) latency/delay is an important non-functional requirement for such services. For example, the minimum E2E latency requirement for pre-sense crash warning is 20-50 ms \cite{traffic_safety1,traffic_safety2}.\\
\indent To that end, mobile edge computing (MEC), the concept of providing computing and storage capabilities near sensors and mobile devices, has been proposed as a viable solution to reduce the E2E latency/delay. MEC has been discussed in the context of ITSs and V2X communication by hosting V2X services and applications on servers near the end-users to reduce the serving latency. As a result, providers can offer new services and applications.\\
\indent Nonetheless, deploying a distributed computing environment consisting of core and edge nodes introduces fresh challenges \cite{AM_ITSM,Li_V2X}. This includes the V2X services/applications' placement decisions. This is because of the limited computing and storage resources available at edge nodes that allow for lower latencies when hosting such services on them. Another challenge is the cost associated with hosting such services on edge nodes, an issue that is of particular interest to service providers. This is due to the higher operational expenditures needed to maintain the edge nodes as compared to cloud nodes because of the larger number of physical structures to deploy and maintain. Therefore, this work focuses on formulating an optimization model that reduces the cost of hosting V2X services while guaranteeing delay, computational, and placement requirements.\\
\indent The contribution of this work is modeling the cost-optimal V2X service placement problem in a distributed cloud/edge computing environment. This is formulated as a binary integer programming model. To the best of our knowledge, this is the first work that considers the problem of cost-optimal V2X service placement while satisfying the computational capabilities at the distributed computing nodes.\\
\indent The organization of this paper is as follows: Section \ref{related_work} briefly discusses a few previous research work in this area. Section \ref{sys_model} describes the system model. Section \ref{prob_formulation} formulates the optimization problem. Section \ref{davsp} presents the proposed heuristic. Section \ref{performance_eval} evaluates the performance and discusses the corresponding results. Section \ref{conc} concludes the paper.
\section{Related Work}\label{related_work}
\indent Few previous works in the literature have investigated the concept of MEC within the context of V2X communications.  A hierarchical MEC architecture was described for adaptive video streaming in which the MEC platform changed the video stream quality according to the observed channel conditions \cite{edge_for_v2x3}. Similarly, MEC technology was proposed to count and classify vehicles for real-time traffic surveillance \cite{edge_for_v2x1}. Similarly, few previous literature works have studied the cost of service placement in MEC environment. For example, a novel virtual network function placement strategy that aims at minimizing a requirement, such as the end to end delay or the overall deployment cost was proposed in \cite{MEC_cost}. However, these works have several limitations as they mainly consider one V2X service or application at a time and investigate the impact of MEC on metrics such as latency while discarding other metrics such as deployment cost and computational capabilities of edge nodes. Although the previous work in \cite{V2X_AM} did study the problem of V2X service placement in such computing environments, but it only considered the delay requirement without studying the associated cost.\\ 
\indent To the best of our knowledge, the problem of cost-optimal placement of V2X services while concurrently considering the availability of computational resources and delay requirements of such services/applications has not been previously investigated. Accordingly, this work formulates the cost-optimal V2X service placement problem in a distributed computing paradigm consisting of core and edge nodes while respecting other non-functional requirements, especially performance and delay of response time. We evaluate the performance of such a system and study the trade-off between the delay and cost of such a placement.
\begin{figure}[!htb]
	\centering
	\includegraphics[scale=.5,trim=0.6cm 0cm 0cm 0cm]{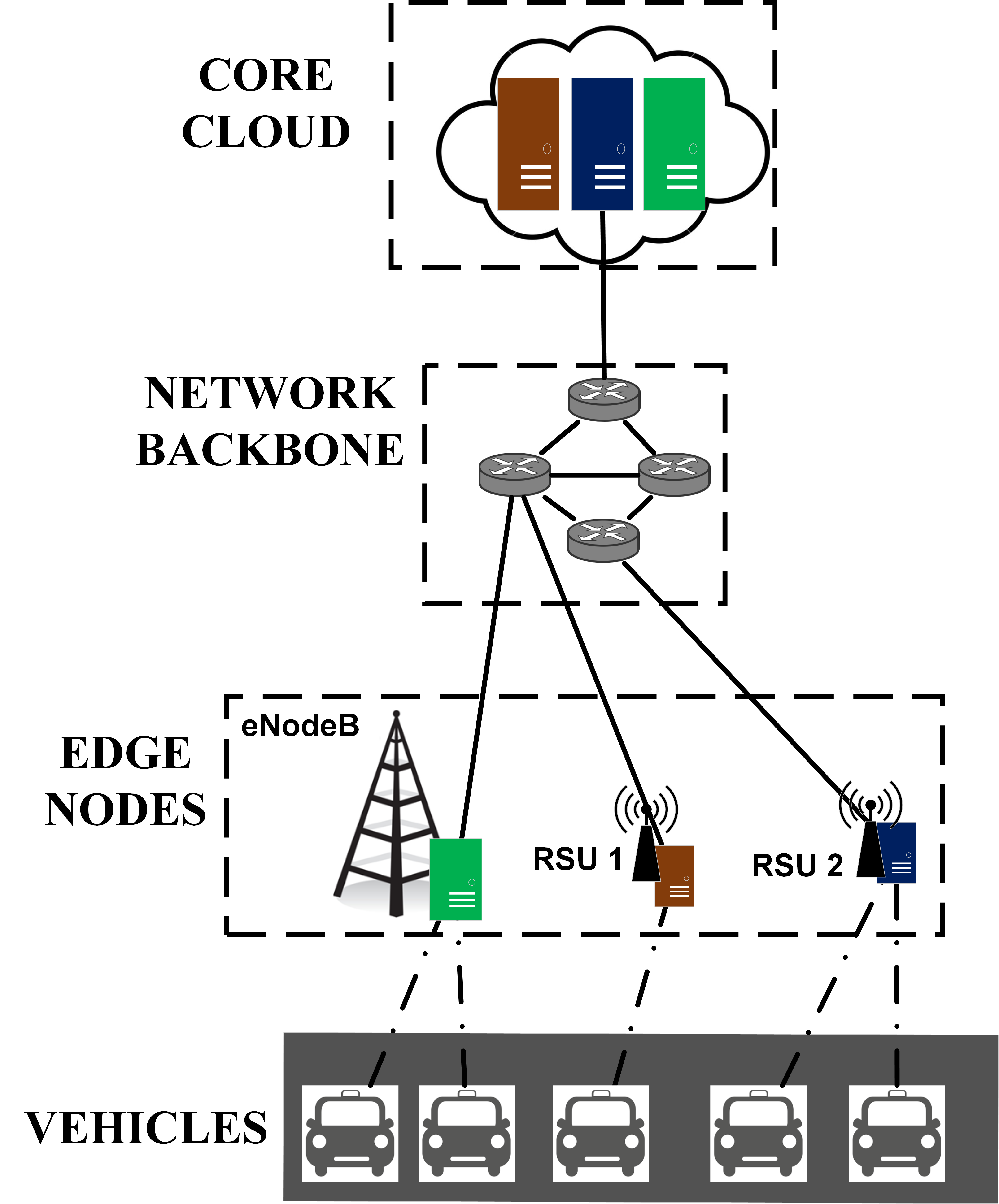}
	\caption{System Model}
	\label{system_model_image}
\end{figure}
\section{System Model}\label{sys_model}
\subsection{System Setup}
\indent  Similar to \cite{V2X_AM}, a uni-directional multiple-lane highway environment is assumed in this work. Furthermore, it is assumed that the vehicles are moving at constant speed with equal distance separation. This mimics vehicles traveling along a highway which on average travel at constant speed. Moreover, this work assumes that the highway segment is covered with LTE-A using multiple evolved NodeB (eNB) base stations. Additionally, multiple road side units (RSUs) are assumed to be placed on the side of the road along the highway. Each eNB or RSU has a MEC host with defined limited computational resources (CPU, memory, and storage). Accordingly, the network access edge is composed of these eNBs and RSUs. Also, it is assumed that the network access edge is connected to a core cloud data center containing servers with larger computing capabilities via the network backbone. This is illustrated in Fig. \ref{system_model_image}. Note that V2X communication is supported by LTE-A through the Uu-based and PC5-based interfaces \cite{CAM_periodicity,AM_aqeeli}.
\subsection{Description of V2X}
\indent Three distinct V2X service types are considered in this work, each representing one use case of different V2X applications. A brief summary of these services and their requirements is given below.
\subsubsection{Cooperative Awareness Basic Service}\mbox{}\\
\indent It is defined by the Cooperative Awareness Message (CAM). The CAM message is routinely exchanged between vehicles and network road nodes. It contains information about the vehicles position, movement, and other sensor data \cite{CAM1,Shaer_V2X}. This service has a strict  latency requirement given that it is one use case of traffic safety services (between 10-20 ms) \cite{messaging_requirement}.    
\subsubsection{Decentralized Environmental Notification Basic Service}
\indent It is defined by the Decentralized Environmental Notification Message (DENM) \cite{DENM1}. The DENM message is a notification message alerting road users of a road-related event (e.g. traffic condition warning). Due to its nature, it is considered a hybrid use case as it belongs to the group of V2X traffic efficiency applications as well as to the group of V2X traffic safety applications. Therefore, it has a less strict latency requirement (tolerates up to 100 ms latency) \cite{messaging_requirement}.
\subsubsection{Media Downloading and Streaming}\mbox{}\\
\indent It is one of the V2X Infotainment applications' use cases \cite{media_streaming1}. Such a service provides vehicle users with on-demand information or entertainment through the Internet \cite{media_streaming1,AM_DNS}. Accordingly, this service has the least strict latency requirement (tolerates latencies up to 1 sec) \cite{media_requirement1,media_requirement2}.
\section{Cost-optimal V2X Service Placement (CO-VSP)}\label{prob_formulation}
\indent This work formulates the cost-optimal placement problem of V2X services in a distributed cloud/edge environment. The goal is to place a set of V2X services in such a manner that minimizes the cost of placing these services on the available computing nodes while adhering to different delay, available computational resources, placement, and cost constraints. This is shown in the CO-VSP model described below. 
\subsection{Key Mathematical Notations:}
\indent Table \ref{math_notations_table} presents the key mathematical notations of this work.
Moreover, the decision variable is:
\begin{equation}
X_{s}^{c} = \begin{cases}
1, & \text{if V2X service/application $s$ is placed on}\\ 
& \text{computing node $c$.}\\
0, & \text{otherwise}.
\end{cases}
\end{equation}

\begin{table}[!ht]
	\centering
	\caption{Key Mathematical Notations}
	\label{math_notations_table}
	\scalebox{1}{
		\begin{tabular}{|p{0.8cm}|p{6.5cm}|}
			\hline
			\textbf{Set} & \textbf{Description} \\ \hline
			$S$ & Set of V2X services' instances to be placed\\ \hline
			$U$ & Set of unique V2X services\\ \hline
			$S_u$ & Subset of V2X services of type $u \in U$\\ \hline
			$C$ & Set of computing nodes (core or edge node) available to host the V2X services\\ \hline
			$V$ & Set of vehicles accessing V2X services' instances\\ \hline
			$d_{s,v}^{c}$ & Delay experienced by vehicle $v$ served by V2X service instance $s$ if placed at computing node $c$\\ \hline
			$D_{s}^{th}$ & Delay/latency threshold of V2X service $s$\\ \hline
			$R_{s}^{i}$ & Computational resource requirement of V2X service $s$ ($i \in \{CPU, memory, storage\}$)\\ \hline
			$Cap_{c}^{i}$ & Computational resource $i$ available at computing node $c$\\ \hline
			$cost_{s}^{c}$ & Cost of placing V2X service $s$ at computing node $c$\\ \hline
			$cost^{th}$ & Cost threshold allocated by the service provider\\ \hline
	\end{tabular}}
\end{table}
\subsection{Problem Formulation:}
\indent The formulation of the cost-optimal V2X service placement (CO-VSP) problem in a distributed computing environment made up of core and edge nodes is:
\subsubsection{Objective Function}
\begin{subequations}\label{problem}
	\begin{itemize}
		\item Equation (\ref{objfunc}) aims to minimize the aggregate cost of placing the V2X services instances. The aggregate cost is determined by summing up all the costs associated with placing a service $s$ at computing node $c$. When the decision variable is set to 1, only the associated cost will be computed as part of the aggregate cost.
		\begingroup\makeatletter\def\f@size{10}\check@mathfonts
		\begin{equation} \label{objfunc}
		\min \sum\limits_{s \in S}\sum\limits_{c \in C} X_{s}^{c} *cost_{s}^{c}
		\end{equation}
		\endgroup
	\end{itemize}
	\subsubsection{Constraints}
	\begin{itemize}		
		\item \textbf{Delay Constraint:} Equation (\ref{delay_constraint}) ensures that the maximum delay/latency threshold of the service is satisfied.
		\begingroup\makeatletter\def\f@size{10}\check@mathfonts
		\begin{equation}\label{delay_constraint}
		\sum\limits_{c \in C} X_{s}^{c} \left(\frac{1}{|V|}\sum\limits_{v \in V} d_{s,v}^{c}\right) \leq D_{s}^{th};\; \forall s \in S
		\end{equation}
		\endgroup
		\item \textbf{Computational Resource Availability Constraint:} Equation (\ref{resources_constraint}) ensures that the aggregate computational resource requirement of the placed V2X service instances placed at computing node $c$ does not exceed its available computational resources.
		\begingroup\makeatletter\def\f@size{10}\check@mathfonts
		\begin{equation} \label{resources_constraint}
		\sum\limits_{s \in S}  X_{s}^{c} R_{s}^{i} \leq Cap_{c}^{i};\; \forall c \in C, \; \forall i \in \{CPU, memory, storage\}
		\end{equation}
		\endgroup
		\item \textbf{Placement Constraint:} Equation (\ref{placement_constraint}) guarantees that each V2X service instance $s$ is only placed at one computing node.
		\begingroup\makeatletter\def\f@size{10}\check@mathfonts
		\begin{equation}\label{placement_constraint}
		\sum\limits_{c \in C} X_{s}^{c} =1 ;\; \forall s \in S
		\end{equation}
		\endgroup
		\item \textbf{Unique Service Placement Constraint:} Equation (\ref{redundant_instances_placement_constraint}) guarantees that instances of unique V2X service of type $u$ are hosted in varying nodes for redundancy purposes.
		\begingroup\makeatletter\def\f@size{10}\check@mathfonts
		\begin{equation}\label{redundant_instances_placement_constraint}
		\sum\limits_{s\in S_u} X_{s}^{c}  \leq 1;\; \forall c \in C;\; \forall u \in U
		\end{equation}
		\endgroup
		\item \textbf{Cost Constraint:} Equation (\ref{cost_constraint}) specifies that the aggregate cost of placement is less than the maximum cost threshold allocated by the service provider. 
		\begingroup\makeatletter\def\f@size{10}\check@mathfonts
		\begin{equation}\label{cost_constraint}
		\sum\limits_{s \in S}\sum\limits_{c \in C} X_{s}^{c} *cost_{s}^{c}  \leq cost^{th}; 
		\end{equation}
		\endgroup
	\end{itemize}
\end{subequations}
Note that this work assumes that the number of vehicles is proportional to the number of instances of each service type. Accordingly, the number of vehicles considered can have a direct impact on the cost since additional V2X service instances would need to be deployed and hosted with the increased number of vehicles.
\subsection{Complexity:}
\indent The problem is considered to be NP-complete since it is a binary integer linear programming problem similar to the facility location problem \cite{bip_complexity,bip_complexity2}. This is further emphasized when determining the search space of the problem. The search space for this problem is $2^{|C||S|}$. For example, when $|C|=10$ and $|S|=5$, the size of the search space becomes $1.125 \times 10^{15}$. As such, it becomes computationally expensive to solve such a problem because of the search space's exponential growth. 
Therefore, it is crucial to develop a low-complexity heuristic algorithm as a viable solution.
\section{Delay-Aware V2X Service Placement (DA-VSP) Algorithm}\label{davsp}
\indent The cost metric can be introduced to any general service placement strategy/algorithm. However, we select the placement algorithm of \cite{V2X_AM} to demonstrate our approach. The work presented in this manuscript proposes a low complexity heuristic, namely ``Delay-aware V2X Service Placement'' (DA-VSP) algorithm, as a solution to the V2X service placement problem. In other words, this algorithm is an adaptation of our previous G-VSPA heuristic algorithm proposed in \cite{V2X_AM} to consider the cost of deployment in a placement algorithm. Algorithm \ref{da-vsp} provides the pseudo-code of the DA-VSP algorithm.
\begin{figure}[!t]
	\centering
	\scalebox{1}{
		\begin{minipage}{\columnwidth}
			\begin{algorithm}[H]
				\caption{Delay-Aware V2X Service Placement (DA-VSP)}
				\label{da-vsp}
				\begin{algorithmic}[1]
					\renewcommand{\algorithmicrequire}{\textbf{Input:}}
					\renewcommand{\algorithmicensure}{\textbf{Output:}}
					\REQUIRE $U = \{1,2,..,|U|\}$, $S = \{1,2,S_1,..,|S|\}$\\ 
					$\;\;\;\;\;C = \{1,2,..,|C|\}$, $V = \{1,2,..,|V|\}$\\
					$\;\;\;\;\;Cost_c = \{Cost_1,Cost_2,..,Cost_C\}$\\
					\ENSURE  $X_{s_u}^{c}$, $Agg\;Cost=\sum\limits_{s \in S}\sum\limits_{c \in C} X_{s}^{c} * Cost_c $\\
					\texttt{\\}
					\STATE \textbf{define} $X_{s_u}=\sum\limits_{c \in C} X_{s_u}^{c}$
					\texttt{\\}
					\texttt{\\}
					\STATE\textbf{set} $U_{sort}=Asc\;Sort(U)$
					\texttt{\\}
					\texttt{\\} 
					\FOR{$u \in U_{sort}$}
					\FOR{$s_u \in S_u$}
					\STATE \textbf{define} $d_{s_u}^c=\left(\frac{1}{|V|}\sum\limits_{v \in V} d_{s,v}^{c}\right)$
					\texttt{\\}
					\texttt{\\}
					\WHILE{$X_{s_u}\neq 1$}
					\STATE \textbf{find} $cost_{s_u}^c = \min \limits_{c \; \in \; C} Cost_c $
					\texttt{\\}
					\texttt{\\}
					\IF{$Rem\;Cap_c > R_{s_u} \&\;d_{s_u}^c< D_{s_u}^{th}$}{
						\STATE \textbf{set} $X_{s_u}^{c} = 1$
						\STATE \textbf{update} $Rem\;Cap_c\;=Rem\;Cap_c-R_{s_u}$
						\STATE \textbf{update} $C \; = C \; \backslash \; c$}
					\ELSE{
						\STATE \textbf{update} $cost_{s_u}^c\;=\; \infty$
					}
					\ENDIF
					\ENDWHILE
					\ENDFOR
					\ENDFOR
					\RETURN $X_{s_u}^{c}$, $Agg\; Cost$
				\end{algorithmic}
			\end{algorithm}
	\end{minipage}}
\end{figure}
\subsection{Description:}
\indent The proposed DA-VSP algorithm is organized as follows:
\begin{itemize}
	\item Line 1: A mock variable is defined that determines if a V2X service instance was placed or not.
	\item Line 2: The unique services are sorted in ascending delay tolerance order.
	\item Lines 3-16: Algorithm iterates  through the instances of each unique service type. For each instance to be hosted, the algorithm determines the computing node with the lowest deployment cost (line 7). If this node adheres to the delay constraint as well as the computational resource constraint, the decision variable is set to 1. Moreover, the available computing resources at node $c$ are updated and the node is removed from the set of available nodes for all other instances of the same type $u$ (lines 8-11). If the node violates any of these constraints, the node is ignored and the algorithm checks the following node with the second lowest deployment cost (line 12-14). This process is repeated until all the service instances are placed.  
\end{itemize} 
\subsection{Complexity:} 
\indent The time complexity of the DA-VSP algorithm is $O(|C||S|)$. This is clear given that the algorithm examines $|C|$ potential nodes when trying to place a service instance $s$. Accordingly, using the same values for $|C|$ and $|S|$ as before, the algorithm would need around 50 operations to reach convergence.\\
\indent The low-complexity nature of this algorithm makes it suitable for real-time deployment. Service providers can use it to determine a suitable balance between the anticipated cost and the desired performance by deploying the minimum number of RSUs based on the traffic pattern within a particular area of interest. 
\section{Performance Evaluation}\label{performance_eval}
\indent A MATLAB-based simulator was developed to evaluate the performance of the proposed CO-VSP model in terms of deployment cost, average delay/latency, and computational resource utilization. Note that the CO-VSP model is used for benchmarking purposes with any real-life deployment using the proposed DA-VSP heuristic as the placement algorithm.
\subsection{Simulation Setup:}
\indent An LTE-based V2X system is considered in this work that supports the three aforementioned services. The number of instances of each service is proportional to the number of vehicles. The delay tolerance threshold is set at 20 ms for the CAM service, 50 ms for the DENM service, and 150 ms for the media downloading and streaming service \cite{messaging_requirement,media_requirement1,media_requirement2}. Additionally, CAM service is modeled as a medium sized virtual machine (VM) with low computational requirements due to the local update status of this service \cite{service_comp_req}. Similarly, DENM service is considered to be a large-sized VM \cite{service_comp_req}. This is because it processes more data about the traffic conditions, which results in a higher computational requirement. Lastly, the media downloading service is considered to be an extra-large VM to support and satisfy the different download requests \cite{service_comp_req}.\\
\indent For the physical environment, this work considers a highway scenario consisting of 2 lanes, each 2 km long. Moreover, it is assumed that there are 2 core cloud nodes, 3 eNodeBs, and 5 RSUs. Each core cloud node has 32 cores of CPU, 64 GB of memory, and 240 GB of storage. Similarly, each eNB and RSU has 8 cores of CPU, 16 GB of memory, and 240 GB of storage. Moreover, the delays/latencies are considered to be uniformly distributed with Vehicle-to-RSU delay between 1 ms -10 ms \cite{CAM_periodicity}, Vehicle-to-eNB delay between 20 ms - 40 ms \cite{CAM_periodicity,access_delay_range}, and Vehicle-to-core delay between 60 ms -130 ms \cite{access_delay_range}. The cost of hosting a service instance in the cloud is estimated to be 7.5k\$/month (assuming Amazon's web services are used) \cite{cloud_cost}, while the cost of hosting a service instance at an edge node is estimated to be around 15k\$/month (almost a 100\% increase in cost for edge nodes as compared to core cloud nodes) \cite{BS_cost,RSU_cost}. The maximum threshold cost per month is assumed to be 500k\$/month \cite{cost_budget}. 
\subsection{Results and Discussion:}
\begin{figure}[!t]
	\centering
	\includegraphics[scale=0.11,trim=0cm 4cm 0cm 1cm,clip]{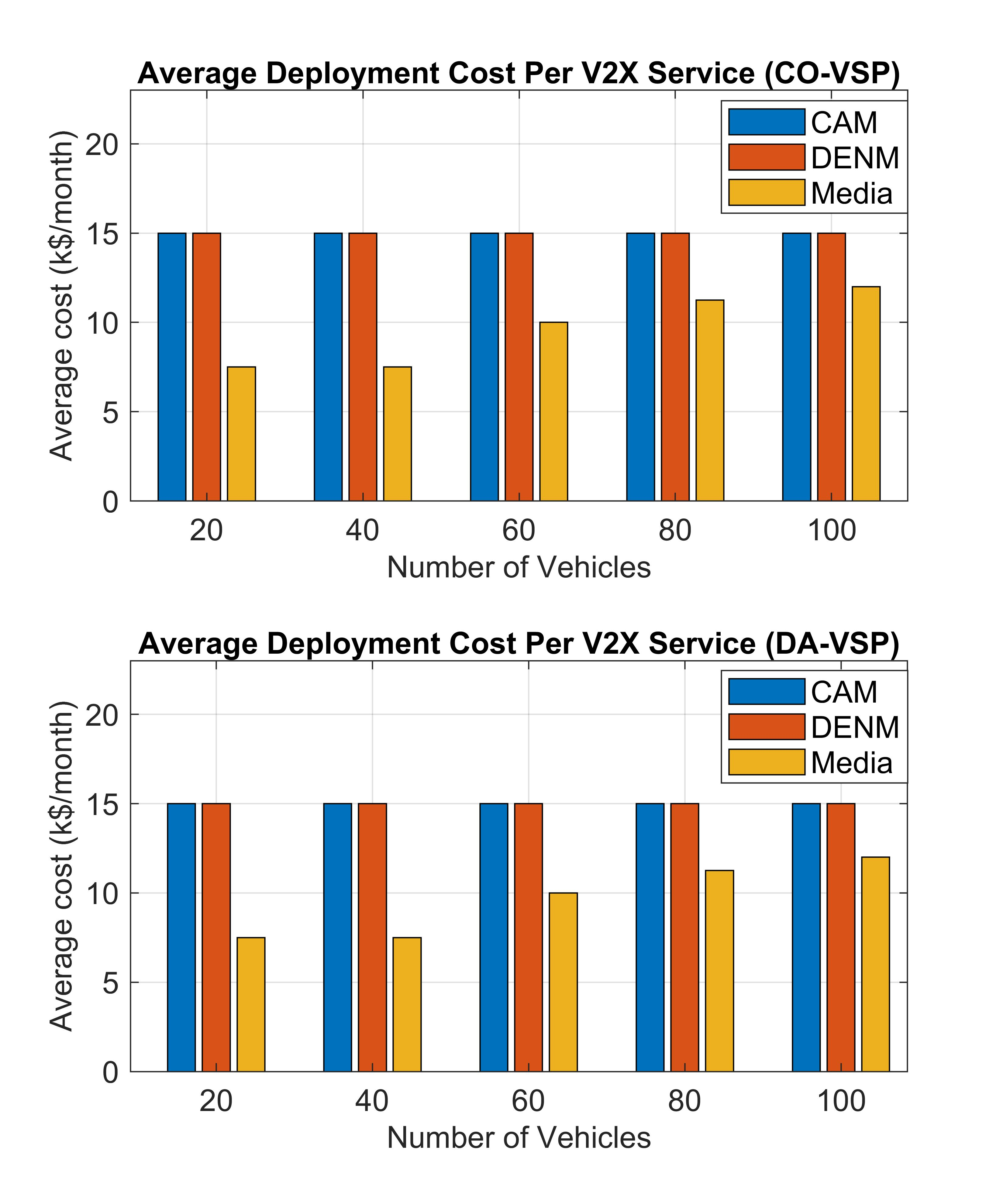}
	\caption{Different V2X Services' Average Monthly Cost}
	\label{cost_fig}
\end{figure}
\begin{figure}[!t]
	\centering
	\includegraphics[scale=0.11,trim=0cm 5cm 0cm 1cm,clip]{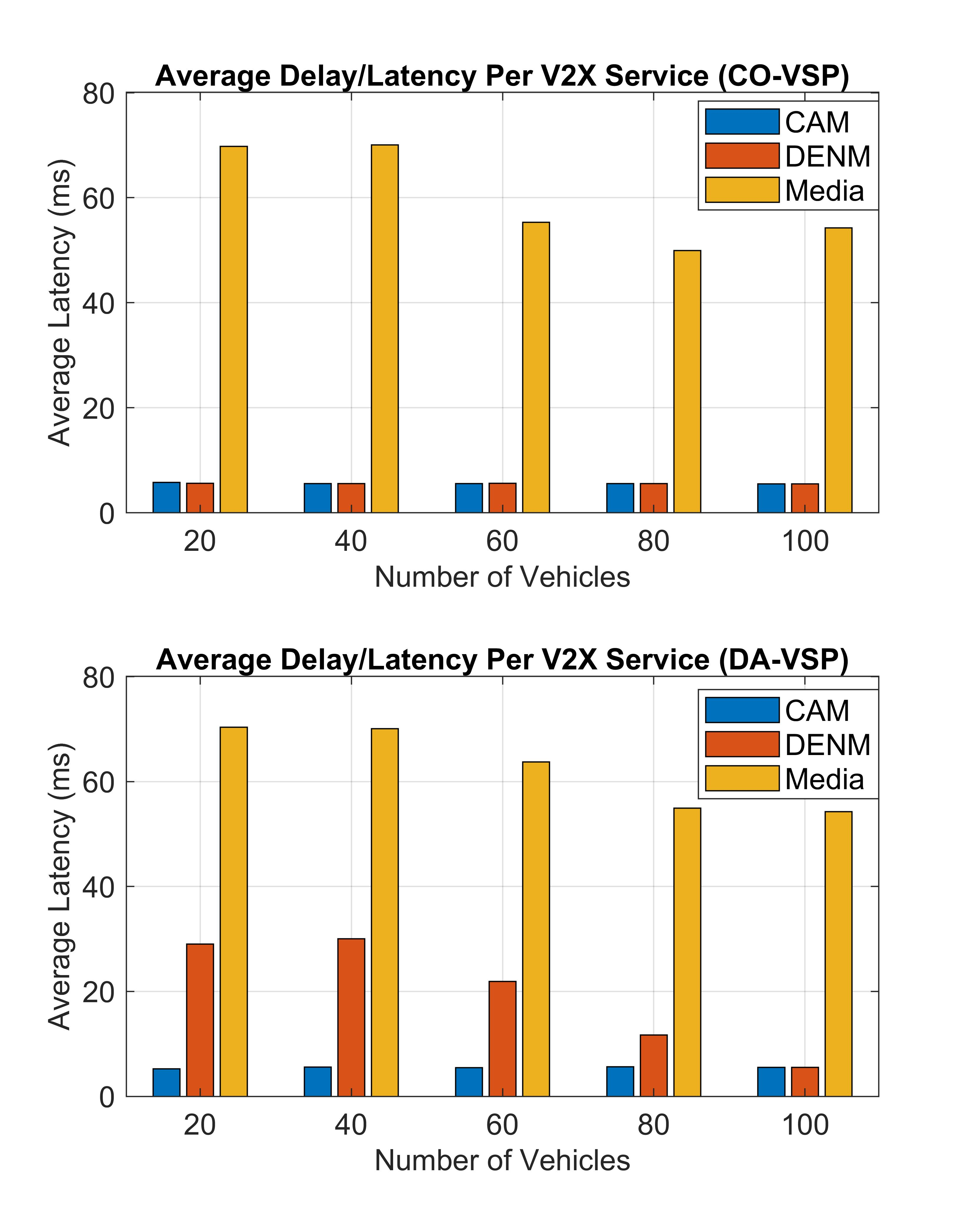}
	\caption{Different V2X Services' Average Delay/Latency}
	\label{delay_fig}
\end{figure}
\begin{figure}[!t]
	\centering
	\includegraphics[scale=0.11,trim=0cm 1cm 0cm 1cm,clip]{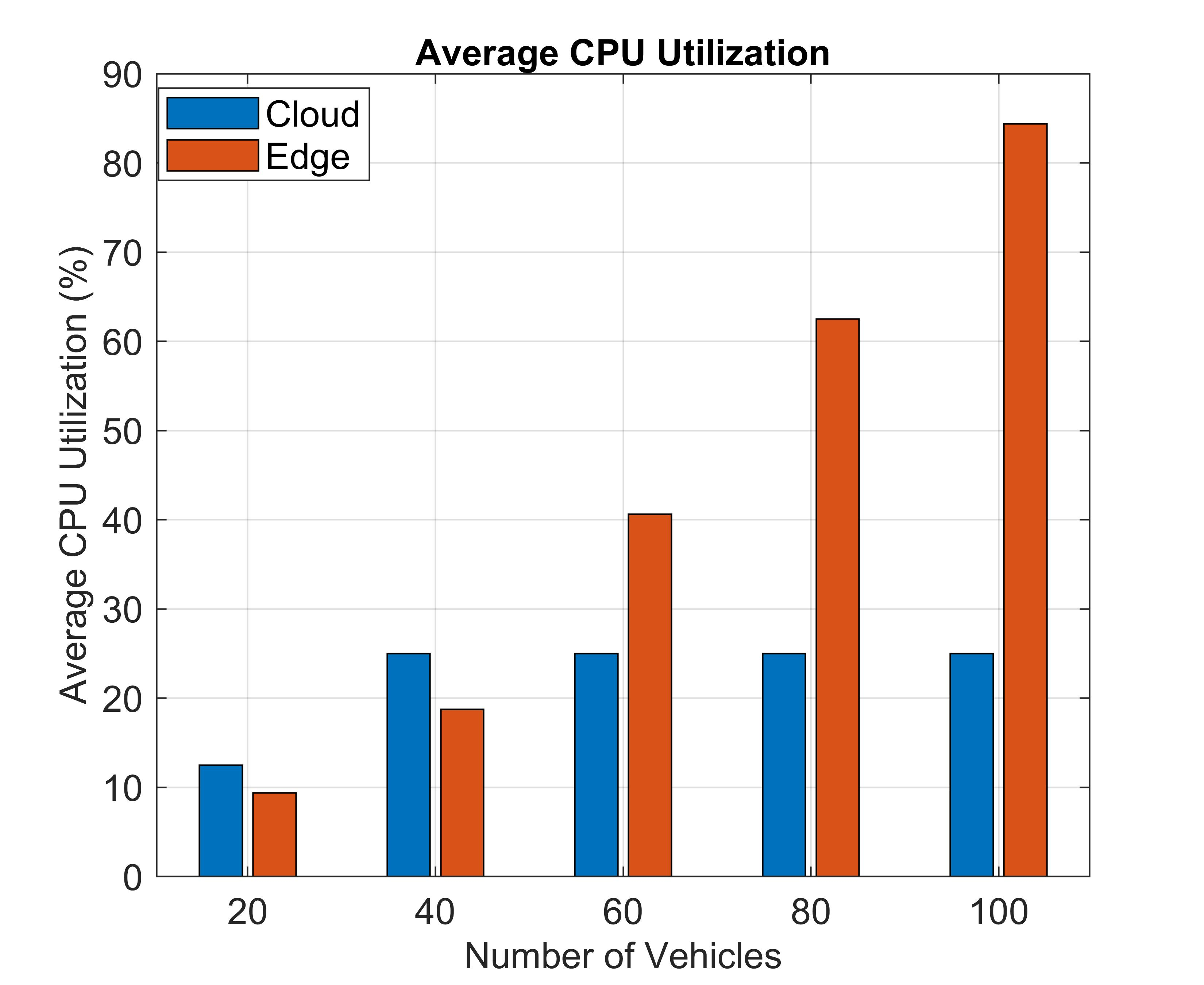}
	\caption{Different Computing Nodes' Average CPU Utilization}
	\label{cpu_fig}
\end{figure}

\indent Fig. \ref{cost_fig} shows the average monthly deployment cost while Fig. \ref{delay_fig} shows the average delay/latency of the V2X services considered. Despite the fact that the proposed CO-VSP model guarantees the QoS requirements of the different V2X service, the trade-off between deployment cost and delay/latency is clearly evident. The average deployment cost for media services increases while its average delay/latency decreases with the increase in number of vehicles. This is expected since there are more media downloading instances being hosted at the edge with the increasing number of vehicles. In contrast, CAM and DENM services remain stable in terms of deployment cost and latency. This is due to the strict requirements in terms of delay for these services which force the model to always place them at the edge rather than the core cloud nodes. The same observations apply when the DA-VSP heuristic algorithm is considered. Moreover, it can be seen that the DA-VSP algorithm achieves similar performance to that of the CO-VSP model while having lower computational complexity. This is because it also tries to find the computing node with the least cost that can satisfy the delay constraint.\\
\indent Fig. \ref{cpu_fig} illustrates the average CPU utilization in both the cloud and edge nodes. As expected, the CPU utilization at the edge increases with the increased number of vehicles. Again, this is because of the greater number of media downloading service instances that move away from the core cloud and are hosted at the edge. In contrast, the cloud utilization becomes stable due to the limited number of instances it can host to satisfy the unique service placement constraint, which forces the model to place some of the delay-tolerant services closer to the user, thus increasing the edge's utilization.\\ 
\indent Moreover, it was shown in \cite{V2X_AM} that the optimal formulation had an average runtime of around 70 ms while the heuristic had an average runtime of around 0.2 ms. Given that the CO-VSP optimization model is similar to the optimization model presented in \cite{V2X_AM} and the fact that the DA-VSP algorithm is an extension to the heuristic proposed in \cite{V2X_AM}, the average runtime for the CO-VSP and DA-VSP are expected to be within the same ranges. This reiterates the real-time deployment potential of the DA-VSP algorithm with the CO-VSP model also being suitable to be used for benchmarking purposes.
\section{Conclusion}\label{conc}
\indent Vehicle-to-everything (V2X) communication, a cornerstone of intelligent transportation systems (ITSs), has garnered more attention from different entities including governmental agencies and automotive manufacturers due to the many projected benefits of such systems. However, offering such services is coupled with maintaining stringent non-functional requirements such as E2E latency and performance. To that end, mobile edge computing (MEC) has been posited as a viable solution to reduce the E2E latency/delay. Nonetheless, adopting a distributed computing environment introduces fresh challenges including the placement decisions of the V2X applications/services and the corresponding cost associated with hosting such services on edge nodes.\\ 
\indent Therefore, this work formulated the cost-optimal V2X service placement (CO-VSP) problem while considering delay, computational, and placement requirements as a binary integer programming model. Moreover, this work developed a low-complexity heuristic titled ``Delay-Aware V2X Service Placement (DA-VSP) Algorithm'' as a solution to this problem. Through extensive simulation, it was shown that the proposed CO-VSP model successfully guaranteed the functional and non-functional requirements of the various V2X services. Moreover, the results highlighted the trade-off between deployment cost and latency in which delay-tolerant services tended to be placed at the cloud core to reduce the cost while delay-stringent services were placed at the edge to maintain their QoS requirements. Furthermore, it was observed that the proposed DA-VSP algorithm had comparable performance to the CO-VSP model while having lower computational complexity.

\section*{Acknowledgments}
This work is partially supported by the Natural Sciences and Engineering Research Council of Canada (NSERC) [NSERC Strategic Partnership Grant STPGP - 521537].

\small
\bibliographystyle{IEEEtran}
\balance
\bibliography{References}

\begin{thebibliography}{10}
\providecommand{\url}[1]{#1}
\csname url@samestyle\endcsname
\providecommand{\newblock}{\relax}
\providecommand{\bibinfo}[2]{#2}
\providecommand{\BIBentrySTDinterwordspacing}{\spaceskip=0pt\relax}
\providecommand{\BIBentryALTinterwordstretchfactor}{4}
\providecommand{\BIBentryALTinterwordspacing}{\spaceskip=\fontdimen2\font plus
\BIBentryALTinterwordstretchfactor\fontdimen3\font minus
  \fontdimen4\font\relax}
\providecommand{\BIBforeignlanguage}[2]{{%
\expandafter\ifx\csname l@#1\endcsname\relax
\typeout{** WARNING: IEEEtran.bst: No hyphenation pattern has been}%
\typeout{** loaded for the language `#1'. Using the pattern for}%
\typeout{** the default language instead.}%
\else
\language=\csname l@#1\endcsname
\fi
#2}}
\providecommand{\BIBdecl}{\relax}
\BIBdecl

\bibitem{v2x_access_survey}
Z.~MacHardy, A.~Khan, K.~Obana, and S.~Iwashina, ``{V2X Access Technologies:
  Regulation, Research, and Remaining Challenges},'' \emph{IEEE Communications
  Surveys Tutorials}, vol.~20, no.~3, pp. 1858--1877, thirdquarter 2018.

\bibitem{v2x_communication_modes}
{3rd Generation Partnership Project (3GPP)}, ``{Technical specification group
  services and system aspects; study on LTE support of Vehicle to Everything
  (V2X) services. Release 14},'' 3GPP, Tech. Rep. TR 22.885, Dec. 2015.

\bibitem{traffic_safety1}
C.~V. S.~C. Consortium \emph{et~al.}, ``Vehicle safety communications project:
  Task 3 final report: identify intelligent vehicle safety applications enabled
  by dsrc,'' \emph{National Highway Traffic Safety Administration, US
  Department of Transportation, Washington DC}, 2005.

\bibitem{traffic_safety2}
ETSI, ``Intelligent transport systems (its); vehicular communications; basic
  set of applications; definitions,'' Tech. Rep. ETSI TR 102 638, Tech. Rep.,
  2009.

\bibitem{AM_ITSM}
A.~Moubayed and A.~Shami, ``Softwarization, virtualization, \& machine learning
  for intelligent \& effective v2x communications,'' \emph{IEEE Intelligent
  Transportation Systems Magazine}, 2020.

\bibitem{Li_V2X}
L.~{Yang}, A.~{Moubayed}, I.~{Hamieh}, and A.~{Shami}, ``Tree-based intelligent
  intrusion detection system in internet of vehicles,'' in \emph{2019 IEEE
  Global Communications Conference (GLOBECOM)}, 2019, pp. 1--6.

\bibitem{edge_for_v2x3}
D.~{Sabella}, N.~{Nikaein}, A.~{Huang}, J.~{Xhembulla}, G.~{Malnati}, and
  S.~{Scarpina}, ``{A Hierarchical MEC Architecture: Experimenting the RAVEN
  Use-Case},'' in \emph{2018 IEEE 87th Vehicular Technology Conference (VTC
  Spring)}, Jun. 2018, pp. 1--5.

\bibitem{edge_for_v2x1}
W.~Balid, H.~Tafish, and H.~H. Refai, ``Intelligent vehicle counting and
  classification sensor for real-time traffic surveillance,'' \emph{IEEE
  Transactions on Intelligent Transportation Systems}, vol.~19, no.~6, pp.
  1784--1794, 2018.

\bibitem{MEC_cost}
A.~{Leivadeas}, M.~{Falkner}, I.~{Lambadaris}, M.~{Ibnkahla}, and G.~{Kesidis},
  ``Balancing delay and cost in virtual network function placement and
  chaining,'' in \emph{2018 4th IEEE Conference on Network Softwarization and
  Workshops (NetSoft)}, Jun. 2018, pp. 433--440.

\bibitem{V2X_AM}
A.~{Moubayed}, A.~{Shami}, P.~{Heidari}, A.~{Larabi}, and R.~{Brunner},
  ``{Edge-enabled V2X Service Placement for Intelligent Transportation
  Systems},'' \emph{IEEE Transactions on Mobile Computing}, pp. 1--1, 2020.

\bibitem{CAM_periodicity}
D.~{Martín-Sacristán}, S.~{Roger}, D.~{Garcia-Roger}, J.~F. {Monserrat},
  A.~{Kousaridas}, P.~{Spapis}, S.~{Ayaz}, and C.~{Zhou}, ``{Evaluation of
  LTE-Advanced connectivity options for the provisioning of V2X services},'' in
  \emph{2018 IEEE Wireless Communications and Networking Conference (WCNC)},
  Apr. 2018, pp. 1--6.

\bibitem{AM_aqeeli}
E.~{Aqeeli}, A.~{Moubayed}, and A.~{Shami}, ``{Towards Intelligent LTE Mobility
  Management through MME Pooling},'' in \emph{2015 IEEE Global Communications
  Conference (GLOBECOM)}, 2015, pp. 1--6.

\bibitem{CAM1}
{ETSI}, ``{Intelligent Transport Systems (ITS) Vehicular Communications: Basic
  Set of Applications - Part 2: Specification of Cooperative Awareness Basic
  Service},'' ETSI, 650 Route des Lucioles F-06921 Sophia Antipolis Cedex -
  FRANCE, Tech. Rep., 2011.

\bibitem{Shaer_V2X}
I.~Shaer, A.~Haque, and A.~Shami, ``{Multi-Component V2X Applications Placement
  in Edge Computing Environment},'' in \emph{2020 IEEE International Conference
  on Communications (ICC)}, 2020, pp. 1--6.

\bibitem{messaging_requirement}
Z.~{Amjad}, A.~{Sikora}, B.~{Hilt}, and J.~{Lauffenburger}, ``{Low Latency V2X
  Applications and Network Requirements: Performance Evaluation},'' in
  \emph{2018 IEEE Intelligent Vehicles Symposium (IV)}, Jun. 2018, pp.
  220--225.

\bibitem{DENM1}
{ETSI}, ``{Intelligent Transport Systems (ITS) Vehicular Communications: Basic
  Set of Applications - Part 3: Specifications of Decentralized Environmental
  Notification Basic Service},'' ETSI, 650 Route des Lucioles F-06921 Sophia
  Antipolis Cedex - FRANCE, Tech. Rep., 2010.

\bibitem{media_streaming1}
------, ``{Intelligent Transport Systems (ITS) Vehicular Communications: Basic
  Set of Applications - Definitions},'' ETSI, 650 Route des Lucioles F-06921
  Sophia Antipolis Cedex - FRANCE, Tech. Rep., 2009.

\bibitem{AM_DNS}
A.~{Moubayed}, M.~{Injadat}, A.~{Shami}, and H.~{Lutfiyya}, ``{DNS
  Typo-Squatting Domain Detection: A Data Analytics and Machine Learning Based
  Approach},'' in \emph{2018 IEEE Global Communications Conference (GLOBECOM)},
  2018, pp. 1--7.

\bibitem{media_requirement1}
{ETSI}, ``{LTE-Service Requirements for V2X Services},'' ETSI, 650 Route des
  Lucioles F-06921 Sophia Antipolis Cedex - FRANCE, Tech. Rep., 2017.

\bibitem{media_requirement2}
MUX, ``The low latency live streaming landscape in 2019,'' 2019.

\bibitem{bip_complexity}
R.~M. Karp, ``Reducibility among combinatorial problems,'' in \emph{Complexity
  of computer computations}.\hskip 1em plus 0.5em minus 0.4em\relax Springer,
  1972, pp. 85--103.

\bibitem{bip_complexity2}
N.~Megiddo and A.~Tamir, ``On the complexity of locating linear facilities in
  the plane,'' \emph{Operations research letters}, vol.~1, no.~5, pp. 194--197,
  1982.

\bibitem{service_comp_req}
Microsoft, ``How to: Change the size of a windows azure virtual machine,''
  Available at:
  \url{https://docs.microsoft.com/en-us/previous-versions/dynamicsnav-2013/dn168976(v=nav.70)},
  2013.

\bibitem{access_delay_range}
K.~{Lee}, J.~{Kim}, Y.~{Park}, H.~{Wang}, and D.~{Hong}, ``{Latency of
  Cellular-Based V2X: Perspectives on TTI-Proportional Latency and
  TTI-Independent Latency},'' \emph{IEEE Access}, vol.~5, pp. 15\,800--15\,809,
  2017.

\bibitem{cloud_cost}
D.~Dyachuk, ``The true costs of running cloud native infrastructure,''
  Available at:
  \url{https://schd.ws/hosted_files/kccncna17/5f/True%20Costs%20Dyachuk%20%282%29.pdf},
  2017.

\bibitem{BS_cost}
{Federeal Communications Commission (FCC)}, ``A broadband network cost model,''
  FCC, Tech. Rep., May 2010.

\bibitem{RSU_cost}
N.~{Nikookaran}, G.~{Karakostas}, and T.~D. {Todd}, ``Combining capital and
  operating expenditure costs in vehicular roadside unit placement,''
  \emph{IEEE Transactions on Vehicular Technology}, vol.~66, no.~8, pp.
  7317--7331, Aug 2017.

\bibitem{cost_budget}
J.~Koomey, K.~Brill, P.~Turner, J.~Stanley, and B.~Taylor, ``A simple model for
  determining true total cost of ownership for data centers,'' \emph{Uptime
  Institute White Paper, Version}, vol.~2, p. 2007, 2007.

\end{thebibliography}

\end{document}